\documentclass[runningheads]{llncs}
\usepackage[dvipsnames]{xcolor}
\definecolor{red}{HTML}{f94144}
\definecolor{orange1}{HTML}{f3722c}
\definecolor{orange2}{HTML}{f8961e}
\definecolor{yellow}{HTML}{f9c74f}
\definecolor{green1}{HTML}{90be6d}
\definecolor{green2}{HTML}{43aa8b}
\definecolor{blue1}{HTML}{577590}
\definecolor{blue2}{HTML}{197db4}
\colorlet{comment}{blue}

\usepackage{colortbl}
\definecolor{logMove}{HTML}{000000}
\definecolor{modelMove}{HTML}{666666}
\definecolor{synchronousMove}{HTML}{d9d9d9}

\usepackage[rflt]{floatflt}

\usepackage{tikz}
\usetikzlibrary{arrows,backgrounds,calc,decorations.pathmorphing,decorations.pathreplacing,fit,matrix,patterns,petri, positioning, shapes,shapes.multipart,matrix,positioning,shapes.callouts,shapes.arrows,calc, decorations.shapes}

\usepackage{hyperref}
\usepackage{rotating}
\usepackage{makecell}
\usepackage{caption}
\usepackage{subcaption}
\usepackage{csquotes}
\usepackage{enumitem}
\usepackage{graphicx}
\usepackage{mathtools}
\usepackage{booktabs}
\usepackage{makecell}
\usepackage{multirow}
\usepackage{todonotes}
\usepackage{caption}
\usepackage{subcaption}
\usepackage{placeins}
\usepackage{wrapfig}

\usepackage{amsmath}
\usepackage{amssymb}

\usepackage[ruled,lined,vlined]{algorithm2e}

\SetCommentSty{xCommentSty}
\SetAlgoSkip{}

\makeatletter
\newcommand\footnoteref[1]{\protected@xdef\@thefnmark{\ref{#1}}\@footnotemark}
\makeatother

\newcommand{\universeActivities}{\ensuremath{\mathcal{A}}}

\newcommand{\alignment}{\ensuremath{\gamma}}

\newcommand{\labelFunc}{\ensuremath{\lambda}}
\newcommand{\marking}{\ensuremath{m}}
\newcommand{\mset}{\ensuremath{B}}

\newcommand{\petriNet}{\ensuremath{N}}

\newcommand{\places}{\ensuremath{P}}
\newcommand{\pnArcs}{\ensuremath{F}}

\newcommand{\proj}{\ensuremath{\pi}}

\newcommand{\reachableMarkings}{\ensuremath{\mathcal{R}}}

\newcommand{\sequence}{\ensuremath{\sigma}}

\newcommand{\transition}{\ensuremath{t}}
\newcommand{\transitions}{\ensuremath{T}}
\newcommand{\univAct}{\ensuremath{\mathcal{A}}}

\newcommand{\univMSet}{\ensuremath{{\mathcal{B}}}}

\newcommand{\universeOfPetriNets}{\ensuremath{\mathcal{N}}}
\newcommand{\pt}{T{=}(V,\allowbreak E,\allowbreak \lambda,\allowbreak r)}

\def\mi#1{\mathit{#1}}

\newcommand{\cfbox}[2]{%
    \colorlet{currentcolor}{.}%
    {\color{#1}%
    \fbox{\color{currentcolor}#2}}%
}

\newcolumntype{l}{>{\columncolor{logMove}\color{white}}c}
\newcolumntype{s}{>{\columncolor{synchronousMove}}c}
\newcolumntype{m}{>{\columncolor{modelMove}\color{white}}c}
\newcolumntype{i}{>{\color{black}}c}

\begin{document}
\title{Conformance Checking for Trace Fragments Using Infix and Postfix Alignments}
\titlerunning{Conformance Checking for Trace Fragments}
%
\author{Daniel Schuster\inst{1,2} \orcidID{0000-0002-6512-9580} 
\and
Niklas Föcking\inst{1} 
\and
Sebastiaan J. van Zelst\inst{1,2} \orcidID{0000-0003-0415-1036} 
\and
Wil~M.~P.~van~der~Aalst\inst{1,2} \orcidID{0000-0002-0955-6940} 
}

\authorrunning{D. Schuster et al.}
%
\institute{Fraunhofer Institute for Applied Information Technology FIT, Germany\\
\email{\{daniel.schuster,niklas.foecking,sebastiaan.van.zelst\}@fit.fraunhofer.de}
\and
RWTH Aachen University, Aachen, Germany\\
\email{wvdaalst@pads.rwth-aachen.de}}

\maketitle              
\begin{abstract}
Conformance checking deals with collating modeled process behavior with observed process behavior recorded in event data.
Alignments are a state-of-the-art technique to detect, localize, and quantify deviations in process executions, i.e., traces, compared to reference process models.
Alignments, however, assume complete process executions covering the entire process from start to finish or prefixes of process executions.
This paper defines infix/postfix alignments, proposes approaches to their computation, and evaluates them using real-life event data. 
\keywords{Process mining \and Conformance checking \and Alignments.}
\end{abstract}

\section{Introduction}

Information systems track the execution of organizations' operational processes in detail.
The generated \emph{event data} describe process executions, i.e., \emph{traces}.
\emph{Conformance checking}~\cite{conformance_checking_book} compares traces from event data with process models.
\emph{Alignments} \cite{replaying_history_on_process_models}, a state-of-the-art conformance checking technique, are widely used, e.g., for quantifying process compliance and evaluating process models.

Most conformance checking techniques relate complete traces, covering the process from start to finish, to reference process models.
Processes are often divided into stages representing different logical/temporal phases; thus, conformance requirements can vary by stage. 
Conformance checking for trace fragments covering conformance-critical phases is therefore useful.
Also, event data often needs to be combined from various data sources to analyze a process holistically.
Thus, conformance checking for trace fragments is valuable 
as complete traces are not required.
While there is the notion of prefix alignments~\cite{Adriansyah.phd}, definitions and calculation methods for infix/postfix alignments do not yet exist. 

\begin{figure}[tb]
    \centering
    \includegraphics[clip, trim=0cm 4.2cm 3.2cm 0.0cm, width=.75\textwidth]{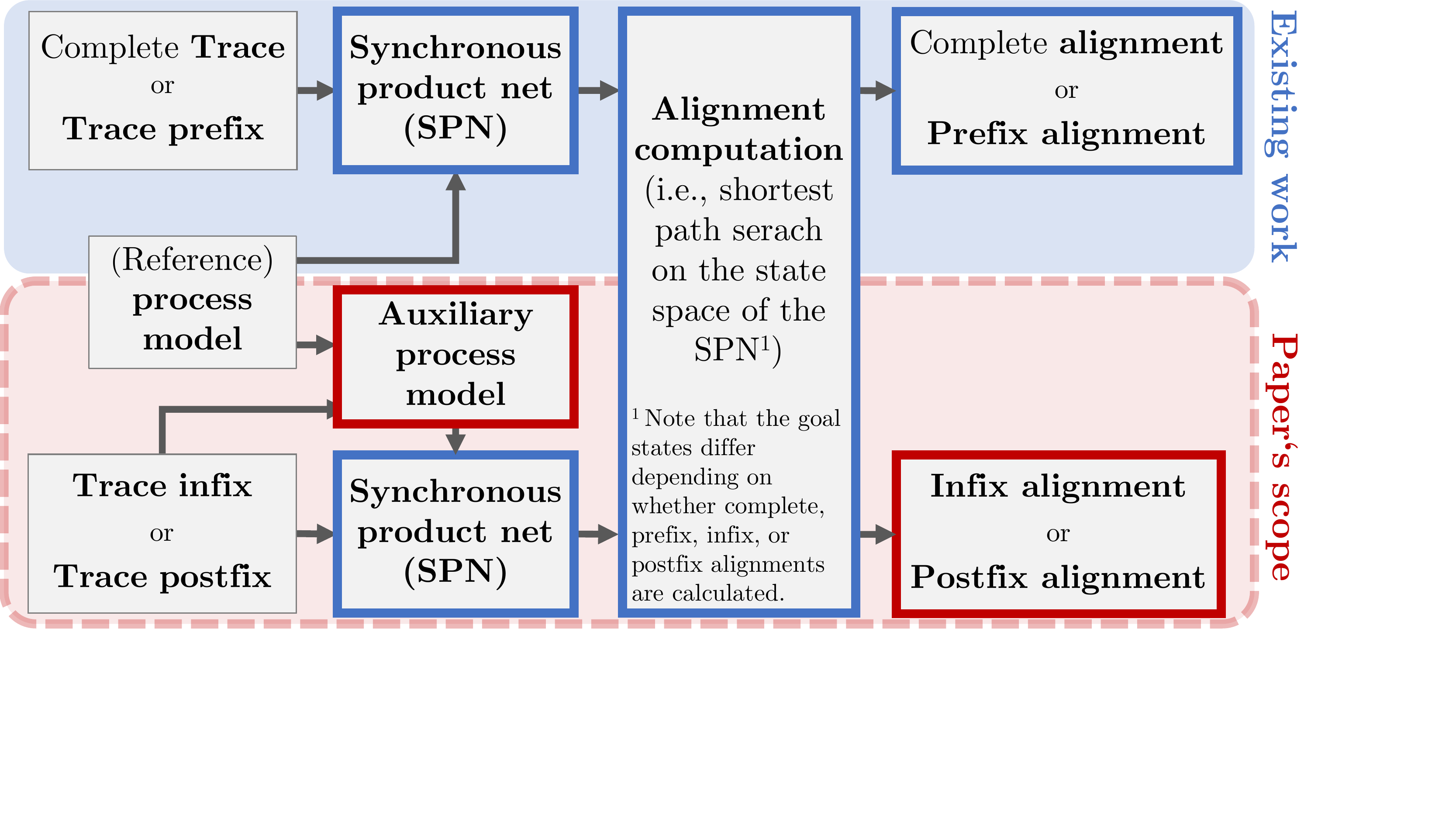}
    \caption{Overview of our approach regarding infix/postfix alignment computation}
    \label{fig:contribution}
\end{figure}

This paper defines infix/postfix alignments and presents their computation. \autoref{fig:contribution} outlines our contributions.
The computation of infix/postfix alignments builds on existing work on calculating (prefix) alignments~\cite{Adriansyah.phd}.
For (prefix) alignment computation, the \emph{synchronous product net (SPN)}~\cite{Adriansyah.phd} is created that defines the search space of the corresponding alignment computation, i.e., a shortest path search.
In this paper, we modify the SPN to adapt it for infix/postfix alignment computation by using an \emph{auxiliary process model} (cf. \autoref{fig:contribution}) as input instead of the reference process model.
The actual search for the shortest path in the state space of the SPN remains unchanged compared to (prefix) alignments apart from different goal states.
We propose two approaches to derive an auxiliary process model. 
One assumes sound workflow nets~\cite{vanderAalst.The_Application_of_Petri_Nets_to_Workflow_Management}, i.e., a subclass of Petri nets often used to model business processes, and the second assumes block-structured workflow nets, i.e., process trees, a subclass of sound WF-nets.


In the remainder of this paper, we present related work (\autoref{sec:related_work}), preliminaries (\autoref{sec:background}), define infix/postfix alignments (\autoref{sec:infix alignment-def}), present their computation (\autoref{sec:infix alignment-comp}), and evaluate the proposed computation (\autoref{sec:evaluation}).

\section{Related Work}
\label{sec:related_work}

We refer to~\cite{conformance_checking_book,Dunzer.conformance_checking_literature_review} for overviews on conformance checking. 
Subsequently, we focus on alignments~\cite{Adriansyah.phd,replaying_history_on_process_models}, which provide a closest match between a trace and a valid execution of a given process model.
In~\cite{Adriansyah.phd,conformance_checking_book} it is shown that alignment computation can be reduced to a shortest path problem.
Further improvements by using alternative heuristics during the search are proposed in~\cite{vanDongen.Efficiently_Computing_Alignments}.
However, the state space of the shortest path problem can grow exponentially depending on the model and the trace~\cite{conformance_checking_book}.
Therefore, approaches for approximating alignments exist, for example, divide-and-conquer~\cite{Taymouri.A_Recursive_Paradigm_for_Aligning_Observed_Behavior} and search space reduction approaches~\cite{vanDongen.Aligning_Modeled_and_Observed_Behavior_A_Compromise_Between_Computation_Complexity_and_Quality}.

Alignments~\cite{Adriansyah.phd,replaying_history_on_process_models} are defined for complete traces that are aligned to a complete execution of a given process model.
Additionally, prefix alignments exist~\cite{Adriansyah.phd}, which are, for example, utilized for online conformance checking~\cite{Schuster.Incremental_State-Space_Expansion}.
In this paper, we define infix/postfix alignments and demonstrate their computation. 
To the best of our knowledge, no related work exists on infix/postfix alignments.

\section{Background}
\label{sec:background}


Given a set $X$, a multiset $\mset$ over $X$ can contain elements of $X$ multiple times. 
For $X {=}\{x,y,z\}$, the multiset $[x^5,y]$ contains $5$ times $x$, once $y$ and no $z$.
The set of all possible multisets over a base set $X$ is denoted by $\univMSet(X)$.
We write $x {\in} B$ if $x$ is contained at least once in multiset $B$.
Given two multisets $b_1,b_2{\in}\univMSet(X)$, we denote their union by $B_1\uplus B_2$.
Finally, given two sets containing multisets, i.e., $B_1,B_2\subseteq\univMSet(X)$, we define the Cartesian 
by $B_1 {\times} B_2 = \{b_1 {\uplus} b_2 \mid b_1 {\in} B_1 \land b_2 {\in} B_2  \}$.
For example, $\big\{[a^2,b],[c]\big\}{\times}\big\{[d^3]\big\}=\big\{[a^2,b,d^3],[c,d^3]\big\}$.

A sequence $\sequence$ of length $|\sequence|{=}n$ over a set $X$ assigns an element to each index, i.e., $\sequence {\colon} \{1,\dots,n\} {\to} X$.
We write a sequence $\sequence$ as $\langle \sequence(1), \sequence(2), ..., \sequence(|\sequence|)\rangle$.
The set of all potential sequences over set $X$ is denoted by $X^*$.
Given $\sequence{\in}X^*$ and $x{\in}X$, we write $x{\in}\sequence$ if $\exists_{1 {\leq} i {\leq} |\sequence|} \big(\sequence(i){=}x\big)$, e.g., $b{\in}\langle a,b\rangle$. 
 Let $\sequence{\in}X^*$ and let $X'{\subseteq}{X}$.
We recursively define $\sequence_{\downarrow_{X'}}{\in}X'^*$ with:
$\langle\rangle_{\downarrow_{X'}} {=} \langle\rangle$, $(\langle x\rangle {\cdot} \sequence)_{\downarrow_{X'}}{=} \langle x \rangle {\cdot} \sequence_{\downarrow_{X'}}$ if $x{\in}X'$ and $(\langle x\rangle {\cdot} \sequence)_{\downarrow_{X'}}{=} \sequence_{\downarrow_{X'}}$ if $x{\notin}X'$.
For a sequence $\sequence{=}\langle(x^1_1,\dots,x^1_n),\dots,(x_1^m,\allowbreak \dots,x_n^m)\rangle \in (X_1{\times}\dots{\times} X_n)^*$ containing $n$-tuples, we define projection functions $\proj^*_1(\sequence){=}\langle x_1^1,\allowbreak\dots,x_1^m\rangle,\allowbreak\dots,\allowbreak\proj^*_n(\sequence){=}\langle x_n^1,\allowbreak \dots,x_n^m\rangle$. 
For instance, $\proj^*_2 ( \langle  (a,b),\allowbreak (c,d),\allowbreak (c,b) \rangle ) {=} \langle b,d,b \rangle$.



\emph{Event data} describe the execution of business processes. 
An event log can be seen as a multiset of process executions, i.e., traces, of a single business process.
We denote the universe of process activity labels by $\univAct$.
Further, we define a \emph{complete/infix/postfix trace} as a sequence of executed activities, i.e., $\sigma {\in} \univAct^*$.


\subsection{Process Models}

\begin{figure}[tb]
    \centering
    \resizebox{.8\textwidth}{!}{%

    \begin{tikzpicture}[node distance=.9cm,>=stealth',bend angle=20,auto, label distance=-1mm]
        \large
     	\tikzstyle{place}=[circle,thick,draw=black,minimum size=4mm]
      	\tikzstyle{transition}=[thick,draw=black,minimum size=5mm]
      	\tikzstyle{silent}=[rectangle,thick,draw=black,fill=black,minimum size=5mm, text=white]
        
        
        \node [place,tokens=1,label=below:{$p_1$}] (p1) {};
    	\node [transition] (A) [right of = p1, label=$t_1$]  {$a$};    
        \node [place] (p2) [above right of= A,label=below:{$p_2$}] {};
        \node [place] (p3) [below right of= A,label=below:{$p_3$}] {};
        \node [transition] (B) [right of = p2, label=$t_2$]  {$b$};
        \node [transition] (C) [right of = p3, label=$t_3$]  {$c$};
        \node [place] (p4) [right of= B,label=below:{$p_4$}] {};
        \node [place] (p5) [right of= C,label=below:{$p_5$}] {};
    	\node [transition] (D) [below right of= p4, label=$t_4$]  {$d$};
        \node [place] (p6) [right of= D,label=below:{$p_6$}] {};
        
        \node [silent] (t5) [above right of= p6, label=$t_5$]  {$\tau$};

        \node [place] (p7) [above right of= t5,label=below:{$p_7$}] {};
        \node [place] (p8) [below right of= t5,label=below:{$p_8$}] {};
        \node [transition] (E) [right of = p7, label=$t_6$]  {$e$};
        \node [transition] (F) [right of = p8, label=$t_7$]  {$f$};
        \node [place] (p9) [right of= E,label=below:{$p_9$}] {};
        \node [place] (p10) [right of= F,label=below:{$p_{10}$}] {};
        \node [transition] (G) [below of= F, label=above:{$t_8$}]  {$g$};
    	\node [silent] (t9) [below right of= p9, label=$t_9$]  {$\tau$};
    	\node [place] (p11) [right of= t9,label=below:{$p_{11}$}] {};
        \node [transition] (H) [right of= p11, label=$t_{10}$]  {$h$};
        \node [place] (p12) [right of= H,label=below:{$p_{12}$}] {};
        
    	\draw [->] (p1) to (A); 
    	\draw [->] (A) to (p2); 
    	\draw [->] (p2) to (B);
    	\draw [->] (B) to (p4);
    	\draw [->] (C) to (p5);
    	\draw [->] (A) to (p3);
    	\draw [->] (p3) to (C);
    	\draw [->] (p5) to (D);
    	\draw [->] (p4) to (D);
    	\draw [->] (D) to (p6);
    	\draw [->] (p6) to (t5);
    	\draw [->] (p7) to (E);
    	\draw [->] (p8) to (F);
    	\draw [->] (F) to (p10);
    	\draw [->] (E) to (p9);
    	\draw [->] (t5) to (p7);
    	\draw [->] (t5) to (p8);
    	\draw [->,bend right] (p6) to (G);
    	\draw [->] (p9) to (t9);
    	\draw [->] (p10) to (t9);
    	\draw [->] (t9) to (p11);
    	\draw [->,bend right] (G) to (p11);
    	\draw [->] (p11) to (H);
    	\draw [->] (H) to (p12);
 
	\end{tikzpicture}
	}
    \caption{Example Petri net, i.e., a sound WF-net, modeling a process consisting of activities $a,\dots,h$. The initial marking $\marking_i{=}[p_1]$, and the final marking $\marking_f{=}[p_{12}]$.}
    \label{fig:petri_net_example}
\end{figure}
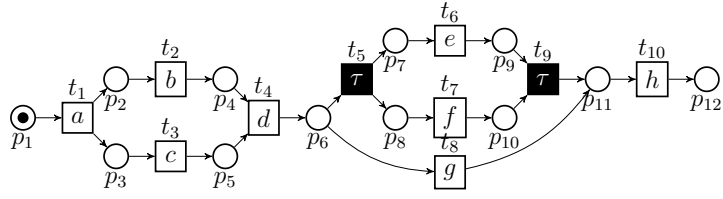

Next, we introduce formalisms to model processes: Petri nets~\cite{vanderAalst.The_Application_of_Petri_Nets_to_Workflow_Management} and process trees. 
\autoref{fig:petri_net_example} shows an example Petri net.
Next, we define accepting Petri nets.

\begin{definition}[Accepting Petri net]
An accepting Petri net $\petriNet=(P,T,F,m_i,\allowbreak m_f,\lambda)$ consists of a finite set of places $P$, a finite set of transitions $T$, a finite set of arcs $F\subseteq (P \times T) \cup (T \times P)$, and a labeling function $\lambda: T \to \universeActivities \cup \{\tau\}$.
We denote the initial marking with $\marking_i \in \univMSet(\places)$ and the final marking with $\marking_f \in \univMSet(\places)$.
\end{definition}

In the remainder of this paper, we say Petri nets when referring to accepting Petri nets.
Given a Petri net $\petriNet{=}(P,T,F,m_i,m_f,\lambda)$ and markings $\marking,\marking'{\in}\univMSet(P)$, if a transition sequence $\sequence{\in}\transitions^*$ leads from $\marking$ to $\marking'$, we write $(\petriNet,\marking) {\xrightarrow{\sequence}} (\petriNet,\marking')$.
If $\marking'$ is reachable from $\marking$,
we write $(\petriNet,\marking) {\rightsquigarrow} (\petriNet,\marking')$.
Further, we write $(\petriNet,\marking)[\transition\rangle$ if $\transition {\in} \transitions$ is enabled in $\marking$.
We let $\reachableMarkings(\petriNet,\marking_i){=}\{\marking'  {\in}\allowbreak \univMSet(\places) \mid (\petriNet,\marking_i) {\rightsquigarrow} (\petriNet,\marking') \}$ denote the state space of $N$, i.e., all markings reachable from $m_i$.
In this paper, we assume that process models are sound workflow nets (WF-nets)~\cite{vanderAalst.The_Application_of_Petri_Nets_to_Workflow_Management}.

\setlength\intextsep{0pt}
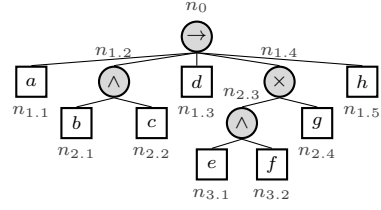
\begin{wrapfigure}{r}{0.45\textwidth}
    \centering
    \scriptsize
    \begin{tikzpicture}[every label/.style={text=darkgray},child anchor=north,parent anchor=south, label distance=.1mm]
        \tikzstyle{tree_op}=[circle,draw=black,fill=gray!30,thick,minimum size=4mm,inner sep=0pt]
        \tikzstyle{tree_leaf}=[rectangle,draw=black,thick,minimum size=4mm,inner sep=0pt]
        \tikzstyle{tree_leaf_inv}=[rectangle,draw=black,fill=black,thick,minimum size=4mm, text=white,inner sep=0pt]
        \tikzstyle{marking}=[dashed, draw=gray]

        \tikzstyle{level 1}=[sibling distance=11mm,level distance=6mm]
        \tikzstyle{level 2}=[sibling distance=10mm,level distance=5.5mm]
        \tikzstyle{level 3}=[sibling distance=8mm, level distance=5.5mm]
        \tikzstyle{level 4}=[sibling distance=10mm,level distance=5mm]
        
        \node [tree_op, label=$n_0$] (root){$\rightarrow $}
            child {node [tree_leaf, label=below:{$n_{1.1}$}] (e) {$a$}}
            child {node [tree_op,label=above:{$n_{1.2}$}] (parallel1) {$\wedge$}
                child {node [tree_leaf, label=below:{$n_{2.1}$}] (c) {$b$}}
                child {node [tree_leaf, label=below:{$n_{2.2}$}] (d) {$c$}}
            }
            child {node [tree_leaf, label=below:{$n_{1.3}$}] (e) {$d$}}
            child {node [tree_op, label=$n_{1.4}$] (loop) {$\times$}
                child { node [tree_op, label=$n_{2.3}$] (choice) {$\wedge$}
                    child {node [tree_leaf, label=below:{$n_{3.1}$}] (a) {$e$}}
                    child {node [tree_leaf, label=below:{$n_{3.2}$}] (f) {$f$}}
                }
                child {node [tree_leaf, label=below:{$n_{2.4}$}] (g) {$g$}}
            }
            child {node [tree_leaf, label=below:{$n_{1.5}$}] (e) {$h$}}
        ;
 
     
    \end{tikzpicture}
    \caption{Process tree $T$ modeling the same process as the WF-net in \autoref{fig:petri_net_example}}
    \label{fig:process_tree_example}
\end{wrapfigure}

Process trees represent \emph{block-structured WF-nets}, a subclass of sound WF-nets~\cite{Leemans.Robust_Process_Mining_with_Guarantees}.
\autoref{fig:process_tree_example} shows an example tree modeling the same behavior as the WF-net in \autoref{fig:petri_net_example}.
Inner nodes represent control flow operators, and leaf nodes represent activities. 
Four operators exist: sequence $(\rightarrow)$, parallel $(\wedge)$, loop $(\circlearrowleft)$, and exclusive-choice $(\times)$.
Next, we define process trees.

\begin{definition}[Process Tree]
\label{def:process-tree-syntax}
Let $\bigoplus {=} \{\rightarrow,\times,\wedge,\circlearrowleft\}$ be the set of operators.
A process tree $T{=}(V,E,\lambda,r)$ consists of a totally ordered set of nodes $V$, a set of edges $E{\subseteq}V{\times}V$, a labeling function $\lambda {:} V {\to} \mathcal{A}{\cup}\{\tau\}{\cup}\bigoplus$, and a root node $r{\in}V$. 

\begin{itemize}[noitemsep,topsep=0pt]
    \item $\big( \{n\},\{\},\lambda,n \big)$ with $\mi{dom}(\lambda){=}\{n\}$ and $\lambda(n){\in} \mathcal{A}{\cup}\{\tau\}$ is a process tree

   \item given $k{>}1$ trees $T_1{=}(V_1,E_1,\lambda_1,r_1),\dots, \allowbreak T_k{=}\allowbreak(V_k,\allowbreak E_k,\allowbreak\lambda_k,r_k)$ with $r{\notin}\allowbreak V_1{\cup}\dots{\cup}V_k$ and $\forall i,j {\in} \{1,\dots,k\}(i{\neq} j \Rightarrow V_i {\cap} V_j {=} \emptyset)$ then $\pt$ is a tree s.t.:
   \begin{itemize}[noitemsep,topsep=0pt]
       \item $V{=}V_1{\cup}\dots{\cup}V_k{\cup}\{r\}$
       \item $E{=}E_1{\cup}\dots{\cup}E_k{\cup}\big\{ (r,r_1),\dots,(r,r_k) \big\}$
       \item $\mi{dom}(\lambda){=}V$ with $\lambda(x){=}\lambda_j(x)$ for all $j{\in}\{1,\dots,k\}, x{\in}V_j$,\\ $\lambda(r){\in}\bigoplus$, and $  \lambda(r){=}{\circlearrowleft}\Rightarrow k{=} 2$
       
   \end{itemize}
   
\end{itemize}

\end{definition}

\noindent $\mathcal{T}$ denotes the universe of process trees.
We refer to~\cite{Leemans.Robust_Process_Mining_with_Guarantees} for a definition of process tree semantics. 
Given $T{=}(V,E,\lambda,r){\in}\mathcal{T}$, the child function $c^T {:} V {\to} V^*$ returns a sequence of child nodes, e.g., $c^T(n_{0}){=}\langle n_{1.1},\dots, n_{1.5}\rangle$, cf. \autoref{fig:process_tree_example}.
The parent function $p^T {:} V {\nrightarrow} V$ returns a node's parent; e.g., $p(n_{2.4}) {=} n_{1.4}$.
For $n{\in} V$, $T(n){\in}\mathcal{T}$ denotes the subtree with root $n$; e.g., $T(n_{2.3})$ denotes the subtree rooted at node $n_{2.3}$ (cf. \autoref{fig:process_tree_example}).
For $T {\in} \mathcal{T}$, we denote its language-equivalent WF-net by $\petriNet^T$.

\subsection{Alignments}
\label{sec:alignments}

This section introduces alignments~\cite{Adriansyah.phd,conformance_checking_book}.
\autoref{fig:classical_alignments_example} shows an example for the WF-net shown in \autoref{fig:petri_net_example} and trace $\sigma {=} \langle d,a,e,h\rangle$.
An alignment's first row, i.e., the trace part, equals the given trace if the skip symbol $\gg$ is ignored. 
The second row, i.e., the model part, equals a sequence of transitions (ignoring $\gg$) leading from the initial to the final marking.
An alignment is composed of moves, for instance, each column in \autoref{fig:classical_alignments_example} represents a move; we distinguish four:
\begin{itemize}[noitemsep,topsep=0pt]
    \item 
    \colorbox{synchronousMove}{\color{black}\textbf{synchronous moves}} indicate a match between the model and the trace,
    \item 
    \colorbox{black}{\color{white}\textbf{log moves}} indicate a \emph{mismatch}, i.e., the current activity in the trace is not replayed in the model,
    \item  
    \colorbox{modelMove}{\color{white}\textbf{visible model moves}} indicate a \emph{mismatch}, i.e., the model executes an activity not observed in the trace at this stage, and
    \item  
    \cfbox{black}{\textbf{invisible model moves}} indicate \emph{no} real mismatch, i.e., a model move on a transition labeled with $\tau$.
\end{itemize}

\begin{figure}[tb]
    \scriptsize
    \centering
        \newcolumntype{l}{>{\columncolor{logMove}\color{white}}c}
        \newcolumntype{s}{>{\columncolor{synchronousMove}}c}
        \newcolumntype{m}{>{\columncolor{modelMove}\color{white}}c}
        \newcolumntype{i}{>{\color{black}}c}
        \begin{tabular}{| l | s | m | m | m | i | s| m |i| s |}\hline
        $d$
        & $a$
        & $\gg$
        & $\gg$
        & $\gg$
        & $\gg$
        & $e$
        & $\gg$
        & $\gg$
        & $h$
        
        \\ \hline\hline
        
        \makecell{$\gg$} 
        & \makecell{$t_1$\\$(\lambda(t_1){=}a)$}
        & \makecell{$t_3$\\$(\lambda(t_3){=}c)$}
        & \makecell{$t_2$\\$(\lambda(t_2){=}b)$}
        & \makecell{$t_4$\\$(\lambda(t_4){=}d)$}
        & \makecell{$t_5$\\ }
        & \makecell{$t_6$\\$(\lambda(t_6){=}e)$}
        & \makecell{$t_7$\\$(\lambda(t_7){=}f)$}
        & \makecell{$t_9$\\ }
        & \makecell{$t_{10}$\\$(\lambda(t_{10}){=}h)$}
    \\ \hline
    \end{tabular}
    
    \caption{Optimal alignment for the WF-net shown in \autoref{fig:petri_net_example} and $\sigma {=} \langle d,a,e,h \rangle$}
    \label{fig:classical_alignments_example}
\end{figure}



\noindent Since we are interested in an alignment finding the closest execution of the model to a given trace, the notion of optimality exists.
An alignment for a model and trace is \emph{optimal} if no other alignment exist with less visible model and log moves.

\section{Infix and Postfix Alignments}
\label{sec:infix alignment-def}

This section defines infix and postfix alignments.
Infix alignments align a given trace infix against an infix of the WF-net's language. 
Thus, the model part of an infix alignment starts at some reachable marking from the given WF-net's initial marking and ends at an arbitrary marking.
\autoref{fig:infix alignments} depicts two infix alignments for the WF-net shown in \autoref{fig:petri_net_example}.
As for alignments, the first row of an infix alignment corresponds to the given trace infix (ignoring $\gg$).
The second row corresponds to a firing sequence (ignoring $\gg$) starting from a WF-net's reachable marking.

\begin{figure}[tb]
    \centering
    \begin{subfigure}[b]{0.49\textwidth}
        \centering
        \scriptsize
        \begin{tabular}{| s | i |s|}\hline
            $d$
            & $\gg$
            & $g$
            \\ \hline\hline
            \makecell{$t_4$\\$(\lambda(t_4){=}d)$}
            & \makecell{$t_5$}
            & \makecell{$t_8$\\$(\lambda(t_8){=}g)$}
            \\ \hline
        \end{tabular}
        \caption{Infix alignment for $\sigma {=} \langle d,g\rangle$}
        \label{fig:infix alignments-a}
    \end{subfigure}
    \begin{subfigure}[b]{0.49\textwidth}
        \centering
        \scriptsize
        \begin{tabular}{| s | s | i |s|}\hline
            $b$
            & $d$
            & $\gg$
            & $f$
            \\ \hline\hline
            
            \makecell{$t_2$\\$(\lambda(t_2){=}b)$}
            & \makecell{$t_4$\\$(\lambda(t_4){=}d)$}
            & \makecell{$t_5$}
            & \makecell{$t_7$\\$(\lambda(t_7){=}f)$}
            \\ \hline
        \end{tabular}
        \caption{Infix alignment for $\sigma {=} \langle b,d,f\rangle$}
    \end{subfigure}
    \begin{subfigure}[b]{0.49\textwidth}
        \centering
        \scriptsize
        \begin{tabular}{| s |s| m |}\hline
            $d$
            & $g$
            & $\gg$
            \\ \hline\hline
            \makecell{$t_4$\\$(\lambda(t_4){=}d)$}
            & \makecell{$t_8$\\$(\lambda(t_8){=}g)$}
            & \makecell{$t_{10}$\\$(\lambda(t_{10}){=}h)$}
            \\ \hline
        \end{tabular}
        \caption{Postfix alignment for $\sigma{=}\langle d,g\rangle$}
    \end{subfigure}
    \begin{subfigure}[b]{0.49\textwidth}
        \centering
        \scriptsize
        \begin{tabular}{| l | s | s | m |}\hline
            $a$
            & $d$
            & $g$
            & $\gg$
            \\ \hline\hline
            
            $\gg$
            & \makecell{$t_4$\\$(\lambda(t_4){=}d)$}
            & \makecell{$t_8$\\$(\lambda(t_8){=}g)$}
            & \makecell{$t_{10}$\\$(\lambda(t_{10}){=}h)$}
            \\ \hline
        \end{tabular}
        \caption{Postfix alignment for $\sigma{=}\langle a,d,g\rangle$}
    \end{subfigure}
    \caption{Optimal infix and postfix alignments for the WF-net shown in \autoref{fig:petri_net_example}}
    \label{fig:infix alignments}
\end{figure}

Postfix alignments follow the same concept as infix alignments.
A postfix alignment's model part starts at a reachable marking but ends at the WF-net's final marking.
\autoref{fig:infix alignments} shows examples of postfix alignments for the WF-net shown in \autoref{fig:petri_net_example}.
As for alignments, the notion of optimality applies equally to infix and postfix alignments.
Next, we define complete, infix, and postfix alignments.

\begin{definition}[Complete/infix/postfix alignment]
\label{def:infix postfix alignment}
Let $\sequence{\in}\univAct^*$ be a complete/infix/postfix trace, $\petriNet{=}(\places,\transitions,\pnArcs,\allowbreak m_i,m_f,\labelFunc)$ be a WF-net, and $\gg {\notin} \univAct{\cup}\transitions$.
A sequence $\alignment{\in}\big((\univAct{\cup}\{\gg\})\times(\transitions{\cup}\{\gg\})\big)^*$ is an complete/infix/postfix alignment if:
\begin{enumerate}[noitemsep,topsep=0pt]
    \item $\sequence {=} \proj^*_1(\alignment)_{\downarrow_{\univAct}}$
    \item  
    \begin{itemize}[noitemsep,topsep=0pt]
        \item \textbf{Complete alignment:}
        $(N,m_i)
        \xrightarrow{\proj^*_2(\alignment)_{\downarrow_{\transitions}}}
        (N,m_f)$
    
        \item \textbf{Infix alignment:}\\
        $(N,m_i)
        \rightsquigarrow
        (N,m_1)
        \xrightarrow{\proj^*_2(\alignment)_{\downarrow_{\transitions}}}
        (N,m_2)
        \rightsquigarrow
        (N,m_f)$ for $m_1,m_2{\in}\reachableMarkings(N,m_i)$
        
        \item \textbf{Postfix alignment:}\\
        $(N,m_i)
        \rightsquigarrow
        (N,m_1)
        \xrightarrow{\proj^*_2(\alignment)_{\downarrow_{\transitions}}}
        (N,m_f)$ for $m_1{\in}\reachableMarkings(N,m_i)$
    \end{itemize}
    
    \item $(\gg,\gg) {\notin} \alignment \ \land \ \forall_{a{\in}\univAct, t{\in}T} \big( \labelFunc(t) {\neq} a \Rightarrow (a,t) {\notin} \alignment \big)$
\end{enumerate}

\end{definition}

\section{Computing infix/postfix alignments}
\label{sec:infix alignment-comp}
The given reference process model cannot be immediately used to compute infix/postfix alignments because it requires starting in the initial marking $m_i$.
Thus, our approach (cf. \autoref{fig:contribution}) constructs an \emph{auxiliary process model}.

Reconsider the second requirement of the infix/postfix alignments definition.
For both infix/postfix alignments, the model part starts with a transition enabled in marking $m_1$ that is reachable from the initial marking $m_i$.
Hereinafter, we refer to candidate markings for $m_1$ (cf. \autoref{def:infix postfix alignment}) as \emph{relevant markings}.
The central question is how to efficiently calculate relevant markings that might represent the start of an infix/postfix alignment in its model part.
Below, we summarize our overall approach for infix/postfix alignment computation.

\begin{enumerate}[noitemsep,topsep=0pt]

    \item Calculate \emph{relevant markings} in the given WF-net that may represent the start of the infix/postfix alignment in the model part, cf. $m_1$ in \autoref{def:infix postfix alignment}.
    
    \item Create the auxiliary WF-net using the relevant markings (cf. \autoref{def:auxiliary_wf_net}).

    \item Create the SPN using the \emph{auxiliary} WF-net and the given trace infix/postfix.
    
    \item Perform a shortest path search on the SPN's state space with corresponding goal markings, i.e., goal states regarding the shortest path search.
    \begin{itemize}[topsep=0pt]
        \item \textbf{Infix alignment:} goal markings contain the last place of the SPN's trace net part
        \item \textbf{Postfix alignment:} standard final marking of the SPN~\cite{Adriansyah.phd,conformance_checking_book}
    \end{itemize}
    
    \item Infix/postfix alignment post-processing: removal of the invisible model move that results from using the auxiliary WF-net instead of the original WF-net.
\end{enumerate}
The first two steps are essential, i.e., the generation of the auxiliary WF-net. 
The subsequent SPN generation remains unchanged compared to alignments~\cite{Adriansyah.phd,conformance_checking_book}.
Likewise, the shortest path search on the SPN's state space is unchanged compared to alignments; however, the goal marking(s) differ, see above. 
Subsequently, 
we present two approaches for constructing the auxiliary WF-net. 


\subsection{Baseline Approach for Auxiliary WF-net Construction}
\label{sec:baseline}

This section presents a baseline approach for constructing the auxiliary WF-net. 
This approach assumes a sound WF-net $\petriNet {=} (P,T,F,m_i,m_f,\lambda)$ as reference process model.
As sound WF-nets are \emph{bounded}~\cite{soundness_wf_nets}, their state space is finite.
Thus, we can list all reachable markings $\reachableMarkings(\petriNet,m_i) {=} \{m_1,\dots,m_n\}$; the baseline approach considers all reachable markings as relevant markings.
Given $\petriNet$, the baseline approach adds a new place $p_0$, representing also the new initial marking $[p_0]$, and $n$ silent transitions allowing to reach one of the markings $\{m_1,\dots,m_n\}$ from $[p_0]$.
Thus, when constructing the corresponding SPN using the auxiliary WF-net, it is possible from the SPN's initial marking to execute a transition representing an invisible model move that marks the model part at some reachable marking $m_1$ (cf. \autoref{def:infix postfix alignment}). 
\autoref{fig:baseline_approach_example} shows the auxiliary WF-net of the WF-net shown in \autoref{fig:petri_net_example}.
Below we generally define the auxiliary WF-net for a given set of relevant markings.
Note that for the auxiliary WF-net constructed by the baseline approach, the set of relevant markings $\{m_1,\dots, m_n\} = \reachableMarkings(\petriNet,m_i)$.

\begin{figure}[tb]
    \centering
    \large
    \resizebox{.89\textwidth}{!}{%

    \begin{tikzpicture}[node distance=1cm,>=stealth',bend angle=20,auto,label distance=-1mm]
     	\tikzstyle{place}=[circle,thick,draw=black,minimum size=4mm]
      	\tikzstyle{transition}=[thick,draw=black,minimum size=5mm]
      	\tikzstyle{silent}=[rectangle,thick,draw=black,fill=black,minimum size=5mm, text=white]
      	\tikzstyle{silent2}=[rectangle,thick,draw=black,fill=blue,minimum size=5mm, text=white]
      	\tikzstyle{silent3}=[rectangle,thick,draw=black,fill=red,minimum size=5mm, text=white]
        \clip (-2.9,-3.2) rectangle + (15.9,7.4);
        
        \node [place,tokens=0,label=below:{$p_1$}] (p1) {};
    	\node [transition] (A) [right of = p1, label=$t_1$]  {$a$};    
        \node [place] (p2) [above right of= A,label=below:{$p_2$}] {};
        \node [place] (p3) [below right of= A,label=below:{$p_3$}] {};
        \node [transition] (B) [right of = p2, label=$t_2$]  {$b$};
        \node [transition] (C) [right of = p3, label=$t_3$]  {$c$};
        \node [place] (p4) [right of= B,label=below:{$p_4$}] {};
        \node [place] (p5) [right of= C,label=below:{$p_5$}] {};
    	\node [transition] (D) [below right of= p4, label=$t_4$]  {$d$};
        \node [place] (p6) [right of= D,label=below:{$p_6$}] {};
        
        \node [silent] (t5) [above right of= p6, label=$t_5$]  {$\tau$};

        \node [place] (p7) [above right of= t5,label=below:{$p_7$}] {};
        \node [place] (p8) [below right of= t5,label=below:{$p_8$}] {};
        \node [transition] (E) [right of = p7, label=$t_6$]  {$e$};
        \node [transition] (F) [right of = p8, label=$t_7$]  {$f$};
        \node [place] (p9) [right of= E,label=below:{$p_9$}] {};
        \node [place] (p10) [right of= F,label=below:{$p_{10}$}] {};
        \node [transition] (G) [below of= F, label=below:{$t_8$}]  {$g$};
    	\node [silent] (t9) [below right of= p9, label=$t_9$]  {$\tau$};
    	\node [place] (p11) [right of= t9,label=below:{$p_{11}$}] {};
        \node [transition] (H) [right of= p11, label=$t_{10}$]  {$h$};
        \node [place] (p12) [right of= H,label=below:{$p_{12}$}] {};
        
    	\draw [->] (p1) to (A); 
    	\draw [->] (A) to (p2); 
    	\draw [->] (p2) to (B);
    	\draw [->] (B) to (p4);
    	\draw [->] (C) to (p5);
    	\draw [->] (A) to (p3);
    	\draw [->] (p3) to (C);
    	\draw [->] (p5) to (D);
    	\draw [->] (p4) to (D);
    	\draw [->] (D) to (p6);
    	\draw [->] (p6) to (t5);
    	\draw [->] (p7) to (E);
    	\draw [->] (p8) to (F);
    	\draw [->] (F) to (p10);
    	\draw [->] (E) to (p9);
    	\draw [->] (t5) to (p7);
    	\draw [->] (t5) to (p8);
    	\draw [->,bend right] (p6) to (G);
    	\draw [->] (p9) to (t9);
    	\draw [->] (p10) to (t9);
    	\draw [->] (t9) to (p11);
    	\draw [->,bend right] (G) to (p11);
    	\draw [->] (p11) to (H);
    	\draw [->] (H) to (p12);
    	
    	\node [silent3] (t1_) [left of= p1, label=$t_1'$]  {$\tau$};
    	\node [place,tokens=1,left of=t1_,label=left:{$p_0'$},draw=blue] (p0) {};
    	\draw [->,red] (p0) to (t1_);
    	\draw [->,red] (t1_) to (p1);
    	
    	\node [silent2] (t2_) [above of= A, label=$t_2'$]  {$\tau$};
    	\draw[->,blue] (t2_) to (p2);
    	\draw[->,blue] (t2_) to (p3);
    	\draw[->,blue,bend left=25] (p0) to (t2_);
    	
    	\node [silent2] (t3_) [below of= A, label=$t_3'$]  {$\tau$};
    	\draw[->,blue,bend left=5] (t3_) to (p2);
    	\draw[->,blue,bend right=28] (t3_) to (p5);
    	\draw[->,blue,bend right] (p0) to (t3_);
    	
    	\node [silent3] (t4_) [above of= B, label=$t_4'$]  {$\tau$};
    	\draw[->,red] (t4_) to (p4);
    	\draw[->,red] (t4_) to (p3);
    	\draw[->,red, bend left=25] (p0) to (t4_);
    	
    	\node [silent2] (t5_) [below of= C, label=$t_5'$]  {$\tau$};
    	\draw[->,blue] (t5_) to (p4);
    	\draw[->,blue] (t5_) to (p5);
    	\draw[->,blue, bend right=20] (p0) to (t5_);
    	
    	\node [silent3] (t6_) [below=1.2cm of D, label=$t_6'$]  {$\tau$};
    	\draw[->,red] (t6_) to (p6);
    	\draw[->,red, bend right=30] (p0) to (t6_);
    	
    	\node [silent2] (t7_) [above=1.2cm of t5, label=$t_7'$]  {$\tau$};
    	\draw[->,blue] (t7_) to (p7);
    	\draw[->,blue] (t7_) to (p8);
    	\draw[->,blue, bend left=25] (p0) to (t7_);
    	
    	\node [silent2] (t8_) [above of=E, label=$t_8'$]  {$\tau$};
    	\draw[->,blue] (t8_) to (p9);
    	\draw[->,blue] (t8_) to (p8);
    	\draw[->,blue, bend left=44] (p0) to (t8_);
    	
    	\node [silent3] (t9_) [above of=t8_, label=$t_9'$]  {$\tau$};
    	\draw[->,red] (t9_) to (p7);
    	\draw[->,red,bend left=38] (t9_) to (p10);
    	\draw[->,red, bend left=25] (p0) to (t9_);
    	
    	\node [silent3] (t10_) [below =1.2cm of p10, label=$t_{10}'$]  {$\tau$};
    	\draw[->,red,bend left=28] (t10_) to (p9);
    	\draw[->,red,bend right=40] (t10_) to (p10);
    	\draw[->,red, bend right=25] (p0) to (t10_);
    	
    	\node [silent3] (t11_) [above =1.4cm of p11, label=$t_{11}'$]  {$\tau$};
    	\draw[->,red] (t11_) to (p11);
    	\draw[->,red, bend left=25] (p0) to (t11_);
    	
    	\node [silent2] (t12_) [above =1.9cm of p12, label=$t_{12}'$]  {$\tau$};
    	\draw[->,blue] (t12_) to (p12);
    	\draw[->,blue, bend left=33] (p0) to (t12_);
 
	\end{tikzpicture}
	}
    \caption{Auxiliary WF-net constructed using the baseline approach (\autoref{sec:baseline}) of the WF-net shown in \autoref{fig:petri_net_example}. 
    Red elements are not contained if the baseline approach with subsequent filtering is used (for the example infix $\sigma {=}\langle b,d,f \rangle$).}
    \label{fig:baseline_approach_example}
\end{figure}

\begin{definition}[Auxiliary WF-net]
\label{def:auxiliary_wf_net}
Let $\petriNet {=} (P,T,F,m_i,m_f,\lambda)$ be a WF-net and $\{m_1,\dots,m_n\} {\subseteq} \reachableMarkings(\petriNet,m_i)$ be the given set of relevant markings.
We define the auxiliary WF-net $\petriNet'= (P',T',\allowbreak F',m_i',m_f',\lambda')$ with:
\begin{itemize}[topsep=0pt]
    \item $P' = P \cup \{p_0'\}$ (assuming $p_0'\notin P$)
    \item $T' = T \cup \{t_j' \mid 1 {\leq} j {\leq} n\}$
    \item $F' = F \cup \{ (p_0',t_j') \mid 1 {\leq} j {\leq} n \} \cup \{ (t_j',p) \mid 1 {\leq} j {\leq} n \land p {\in}m_j \}$
    \item $m_i' = [p_0']$ and $m_f' = m_f$
    \item $\lambda'(t_j) {=}\lambda(t_j)$ for all $t_j{\in} T$ and $\lambda'(t_j'){=}\tau$ for all $t_j'{\in} T' {\setminus} T$
\end{itemize}
\end{definition}

When creating the SPN using the auxiliary WF-net and a given trace infix/postfix, the added transitions in the auxiliary WF-net correspond to invisible model moves. 
For example, reconsider the infix alignment in \autoref{fig:infix alignments-a}.
The infix alignment for $\sigma {=} \langle d,g\rangle$ and auxiliary WF-net shown in \autoref{fig:baseline_approach_example} returned after step 4 contains an invisible model move on $t_5'$. 
As this invisible model move on $t_5'$ is the result of using the auxiliary WF-net instead of the original WF-net for which we calculate an infix/postfix alignment, we must remove it, i.e., Step 5.

\subsubsection{Improved Baseline by Subsequent Filtering}
Instead of considering all reachable markings as relevant markings, we filter markings not enabling transitions whose labels are contained in the given infix/postfix $\sigma$.
Reconsider the auxiliary WF-net shown \autoref{fig:baseline_approach_example}; red elements are not included if subsequent filtering is used for the example infix $\sigma {=} \langle b,d,f\rangle$. 
For instance, $t_1'$ is not included, as the marking reached $[p_1]$ only enables $t_1$ with $\lambda(t_1){=}a {\notin} \sigma$. 
Below, we define the relevant markings for a WF-net $N {=} (\places,\transitions,F,m_i,m_f,\lambda)$ and infix/postfix $\sigma$.
$$\big\{ m {\in} \reachableMarkings(N,m_i) \mid \exists_{t \in T} \big( (N,m) [t\rangle \land \lambda(t) {\in} \sequence \big) \big\} \cup \big\{m_f\big\}$$
Note that the auxiliary WF-net constructed by the baseline approach without filtering is independent of the provided trace infix/postfix.
However, the auxiliary WF-net constructed by the baseline plus subsequent filtering depends on the provided model \emph{and} the trace infix/postfix. 

\subsection{Advanced Auxiliary WF-net Construction for Process Trees}
\label{sec:advanced}

This section introduces an advanced approach for constructing an auxiliary WF-net from a given \emph{block-structured} WF-net, i.e., a process tree.
Compared to the baseline, the advanced approach aims to reduce the number of relevant markings.
Further, the advanced approach determines relevant markings directly instead of computing all reachable markings and subsequently filtering (cf. \autoref{sec:baseline}). 

Assume the WF-net from \autoref{fig:petri_net_example} and the infix/postfix $\sigma {=} \langle b,d,f \rangle$.
Reconsider the auxiliary WF-net shown in \autoref{fig:baseline_approach_example}; jumping to marking $[p_2,p_3]$ within the model using the transition $t_{2}'$ does not make sense if we can also jump to marking $[p_2,p_5]$.
From $[p_2,p_3]$ we can replay $b$ and $c$.
However, we need to replay $b$ according to $\sigma$.
Thus, we would always favor the marking $[p_2,p_5]$ over $[p_2,p_3]$ since in the latter one we have to eventually execute $c$ after executing the $b$ to proceed.
Hence, transition $t_{2}'$ allowing to jump to $[p_2,p_3]$ is not needed when computing an optimal infix/postfix alignment for $\langle b,d,f \rangle$. 
The proposed auxiliary WF-net construction in this section is exploiting such conclusions.

\begin{figure}[tb]
    \centering
    \resizebox{.89\textwidth}{!}{%
    \large
    \begin{tikzpicture}[node distance=1cm,>=stealth',bend angle=20,auto,label distance=-1mm]
     	\tikzstyle{place}=[circle,thick,draw=black,minimum size=4mm]
      	\tikzstyle{transition}=[thick,draw=black,minimum size=5mm]
      	\tikzstyle{silent}=[rectangle,thick,draw=black,fill=black,minimum size=5mm, text=white]
      	\tikzstyle{silent2}=[rectangle,thick,draw=black,fill=blue,minimum size=5mm, text=white]
        
        \node [place,tokens=0,label=below:{$p_1$}] (p1) {};
    	\node [transition] (A) [right of = p1, label=$t_1$]  {$a$};    
        \node [place] (p2) [above right of= A,label=below:{$p_2$}] {};
        \node [place] (p3) [below right of= A,label=below:{$p_3$}] {};
        \node [transition] (B) [right of = p2, label=$t_2$]  {$b$};
        \node [transition] (C) [right of = p3, label=$t_3$]  {$c$};
        \node [place] (p4) [right of= B,label=below:{$p_4$}] {};
        \node [place] (p5) [right of= C,label=below:{$p_5$}] {};
    	\node [transition] (D) [below right of= p4, label=$t_4$]  {$d$};
        \node [place] (p6) [right of= D,label=below:{$p_6$}] {};
        
        \node [silent] (t5) [above right of= p6, label=$t_5$]  {$\tau$};

        \node [place] (p7) [above right of= t5,label=below:{$p_7$}] {};
        \node [place] (p8) [below right of= t5,label=below:{$p_8$}] {};
        \node [transition] (E) [right of = p7, label=$t_6$]  {$e$};
        \node [transition] (F) [right of = p8, label=$t_7$]  {$f$};
        \node [place] (p9) [right of= E,label=below:{$p_9$}] {};
        \node [place] (p10) [right of= F,label=below:{$p_{10}$}] {};
        \node [transition] (G) [below of= F, label=below:{$t_8$}]  {$g$};
    	\node [silent] (t9) [below right of= p9, label=$t_9$]  {$\tau$};
    	\node [place] (p11) [right of= t9,label=below:{$p_{11}$}] {};
        \node [transition] (H) [right of= p11, label=$t_{10}$]  {$h$};
        \node [place] (p12) [right of= H,label=below:{$p_{12}$}] {};
        
    	\draw [->] (p1) to (A); 
    	\draw [->] (A) to (p2); 
    	\draw [->] (p2) to (B);
    	\draw [->] (B) to (p4);
    	\draw [->] (C) to (p5);
    	\draw [->] (A) to (p3);
    	\draw [->] (p3) to (C);
    	\draw [->] (p5) to (D);
    	\draw [->] (p4) to (D);
    	\draw [->] (D) to (p6);
    	\draw [->] (p6) to (t5);
    	\draw [->] (p7) to (E);
    	\draw [->] (p8) to (F);
    	\draw [->] (F) to (p10);
    	\draw [->] (E) to (p9);
    	\draw [->] (t5) to (p7);
    	\draw [->] (t5) to (p8);
    	\draw [->,bend right] (p6) to (G);
    	\draw [->] (p9) to (t9);
    	\draw [->] (p10) to (t9);
    	\draw [->] (t9) to (p11);
    	\draw [->,bend right] (G) to (p11);
    	\draw [->] (p11) to (H);
    	\draw [->] (H) to (p12);
    	
    	\node [place,tokens=1,left of=p1,label=left:{$p_0'$},draw=blue] (p0) {};
    	
    	
    	\node [silent2] (t3_) [below of= A, label=$t_3'$]  {$\tau$};
    	\draw[->,blue,bend left=5] (t3_) to (p2);
    	\draw[->,blue,bend right=28] (t3_) to (p5);
    	\draw[->,blue,bend right] (p0) to (t3_);
    	
    	
    	\node [silent2] (t5_) [below of= C, label=$t_5'$]  {$\tau$};
    	\draw[->,blue] (t5_) to (p4);
    	\draw[->,blue] (t5_) to (p5);
    	\draw[->,blue, bend right=25] (p0) to (t5_);
    	
    	
    	
    	\node [silent2] (t8_) [above of=E, label=$t_8'$]  {$\tau$};
    	\draw[->,blue] (t8_) to (p9);
    	\draw[->,blue,bend right=10] (t8_) to (p8);
    	\draw[->,blue, bend left=10] (p0) to (t8_);
    	
    	
    	
    	
    	\node [silent2] (t12_) [above =.35cm of p12, label=$t_{12}'$]  {$\tau$};
    	\draw[->,blue] (t12_) to (p12);
    	\draw[->,blue, bend left=17] (p0) to (t12_);
 
	\end{tikzpicture}
	}
    \caption{Auxiliary WF-net constructed using the advanced approach (cf. \autoref{sec:advanced}) for the block-structured WF-net shown in \autoref{fig:petri_net_example} and the infix $\sigma {=} \langle b,d,f\rangle$}
    \label{fig:advanced_approach_example}
\end{figure}
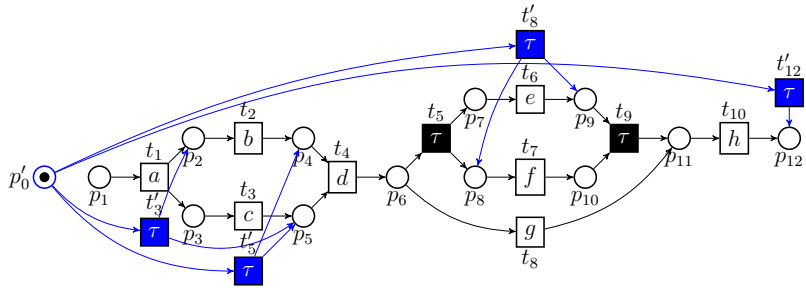

\autoref{fig:advanced_approach_example} shows the auxiliary WF-net that is generated by the advanced approach. 
The shown auxiliary WF-net is specific for the WF-net shown in \autoref{fig:petri_net_example} and the infix/postfix $\sigma=\langle b,d,f\rangle$.
Compared to the auxiliary WF-net generated by the baseline approach (cf. \autoref{fig:baseline_approach_example}), the one shown in \autoref{fig:advanced_approach_example} contains less silent transitions; leading to a reduced state space of the corresponding SPN.
To compute the relevant markings, the advanced approach systematically traverses the given process tree as specified in \autoref{alg:leaf_detection}, which internally calls \autoref{alg:marking_generation} and \autoref{alg:marking_generation_top_down}.

\begin{algorithm}[p]
    \scriptsize
	\caption{Calculating relevant markings for process trees}
	\label{alg:leaf_detection}
	\SetKwInOut{Input}{input}
	\SetKwInOut{Output}{output}
	\Input{${T{=}(V,E,\labelFunc,r)}{\in}\mathcal{T},\
	\sigma{\in}\univAct^*$}
	\Output{$M\subseteq \univMSet(P^T)$} 
	\Begin{
	    \nl $M \gets \{\}$\tcp*[r]{initialize the set of markings for the auxiliary WF-net}
	    \nl let $N^T=(P^T,T^T,F^T,m_i^T,m_f^T,\labelFunc^T)$ be the corresponding WF-net of $T$\;
	    \nl $A\gets \{a \mid a{\in}\univAct \land a{\in}\sigma\}$\tcp*[r]{store all activity labels from $\sigma$ in the set $A$}

	    \ForAll(\tcp*[f]{iterate over leaves whose label is in $\sigma$}){$n\in \{\overline{n}\mid \overline{n}{\in}V \; \land \; \lambda(\overline{n}){\in}A\}$}{
	       
	        \nl $M \gets M\cup \mi{BuMG}\big(T,n,\mi{null},N^T,\emptyset,A\big)$\tcp*[r]{call $\mi{BuMG}$ for each leaf $n$}
	        
        }

		\nl \Return $M \cup \{m_f^T\}$\tcp*[r]{$m_f^T$ is needed for postfix alignments to ensure that the entire model is skippable (i.e., postfix alignment contains log moves only)}\label{line:return_M}
	}			
\end{algorithm}

\begin{algorithm}[p]
	\scriptsize
	\caption{Bottom-up marking generation ($\mi{BuMG}$)}
	\label{alg:marking_generation}
	\SetKwInOut{Input}{input}
	\SetKwInOut{Output}{output}
	\Input{${T{=}(V,E,\labelFunc,r)}{\in}\mathcal{T},\
	n{\in} V,\	
	n'{\in} V,\	
	N^T{=}(P^T,T^T,F^T,m_i^T,m_f^T,\labelFunc^T){\in}\universeOfPetriNets,\ \allowbreak
	M{\subseteq}\univMSet(P^T),\
	A{\subseteq}\univAct$}
	\Output{$M{\subseteq}\univMSet(P^T)$}
	
	\Begin{

	    \nl \uIf(\tcp*[f]{$n$ is a leaf node of $T$}){$\lambda(n)\in\univAct$}{
            \nl let $t\in T^T$ be the transition representing $n\in V$\;
            \nl $M\gets \{[ p \in \bullet t ]\}$\tcp*[r]{initialize $M$ with a marking enabling $t$}\label{line:marking_preset}
        }

        \nl
        \ElseIf(\tcp*[f]{$n$ represents a parallel operator}\label{line:parallel_hit}){$\lambda(n)=\wedge$}{
            
            \nl $S\gets \langle s_1,\dots,s_k \rangle=c^T(n)_{\downarrow_{V{\setminus}\{n'\}}} $\tcp*[r]{$S \in V^*$ contains the siblings of $n'$}
            
            \nl \ForAll{$s_j \in S$}{
                \nl $M_{s_j} \gets \mi{TdMG}\big(T(s_j),N^{T(s_j)},A,\mi{true}\big)$\;\label{line:call_TdMG}
            }
            
            \nl $M \gets M \times M_{s_1} \times \dots \times M_{s_k}$\tcp*[r]{Cartesian product because $\lambda(n)=\wedge$}
            \label{line:cartesian_product}

        }
        \nl \If(\tcp*[f]{node $n$ is the root node of $T$}){$r=n$}{
            \nl \Return $M$\;\label{line:return_M_2}
        }
        \nl $M \gets \mi{BuMG\big(T,{p^T(n)},n,N^T,M,A\big)}$\tcp*[r]{call $\mi{BuMG}$ on $n$'s parent}\label{line:call_BuMG}
	}			
\end{algorithm}

\begin{algorithm}[p]
	\scriptsize
	\caption{Top-down marking generation ($\mi{TdMG}$)}
	\label{alg:marking_generation_top_down}
	\SetKwInOut{Input}{input}
	\SetKwInOut{Output}{output}
	
	\Input{${T{=}(V,E,\labelFunc,r)}{\in}\mathcal{T}, 
	N^T{=}(P^T,T^T,F^T,m_i^T,m_f^T,\labelFunc^T){\in}\universeOfPetriNets, \allowbreak
	A{\subseteq}\univAct, \allowbreak
	\mi{addFinalMarking} {\in} \{\mi{true},\mi{false}\}$}
	
	\Output{$M{\subseteq}\univMSet(P^T)$}
	
	\Begin{
	    \nl \uIf(\tcp*[f]{$r$ is a leaf node})
    {$\lambda(r)\in\univAct$}{
	        \nl let $t\in T^T$ be the transition representing $r$\;
	        \nl $M \gets \emptyset$\;
	        \nl \If{$\lambda(r)\in A$}{
	            \nl $M\gets M \cup  \{ [ p \in \bullet t ] \}$\tcp*[r]{$t$'s label is in the given trace infix/postfix}
	        }
	        \nl \If{$\mi{addFinalMarking}=\mi{true}$}{
	            \nl $M\gets M \cup  \{ [ p \in t\bullet ] \}$\;
	        }
	        \nl \Return $M$\;\label{line:return_final_m}
	    }
	    \nl\Else(\tcp*[f]{$r$ represents an operator}){
	        \nl $S\gets \langle s_1,\dots,s_k \rangle=c^T(r)$\tcp*[r]{$S$ contains all children of the root node $r$}
	        
	        \nl \If{$\lambda(r)=\rightarrow$}{
	            \nl \Return $\mi{TdMG}\big( T(s_1),N^{T(s_1)},A,\mi{false} \big) \cup \dots \cup \mi{TdMG}\big( T(s_{k-1}),\allowbreak N^{T(s_{k-1})},\allowbreak A,\allowbreak \mi{false} \big) \cup \allowbreak 
	            \mi{TdMG}\big( T(s_k),\allowbreak N^{T(s_k)},\allowbreak A,\allowbreak \mi{addFinalMarking} \big)$\;
	        }
	        \nl \If{$\lambda(r)=\wedge$}{
	            \nl \Return $\mi{TdMG}\big( T(s_1),A,N^{T(s_1)},\mi{true} \big) \times\allowbreak \dots \times \allowbreak \mi{TdMG}\big( T(s_k),\allowbreak N^{T(s_k)},\allowbreak A,\allowbreak \mi{true} \big)$\;
	        }
	        \nl \If{$\lambda(r)\in\{\circlearrowleft,\times\}$}{
	            \nl \Return $\mi{TdMG}\big( T(s_1),N^{T(s_1)},A, \mi{addFinalMarking} \big) \cup \mi{TdMG}\big( T(s_{2}),\allowbreak N^{T(s_{2})},\allowbreak A,\allowbreak \mi{false} \big) \cup \dots \cup \mi{TdMG}\big( T(s_k),\allowbreak N^{T(s_k)},\allowbreak A,\allowbreak \mi{false} \big)$\;
	        }
	    } 
	}			
\end{algorithm}

\subsubsection{Restriction to Submodel}
In addition to the described approach, we can further reduce the size of the auxiliary WF-net if we compute \emph{infix alignments}. 
For a process tree $T$, we determine the minimal subtree that contains all leaf nodes whose labels are contained in the given trace infix. 
Since the other subtrees do not contain leaf nodes relevant for the given infix, we can ignore them\footnote{Note that if the determined subtree is placed within a loop, the subtree containing the highest loop and the initial determined subtree has to be considered}. 
Next, we call \autoref{alg:leaf_detection} for the determined subtree and execute the auxiliary WF-net for the determined subtree and the corresponding relevant markings.

\section{Evaluation}
\label{sec:evaluation}

This section presents an evaluation of the infix alignment computation.
We use real-life, publicly available event logs.
We sampled 10,000 infixes per log.
Further, we discovered a process model using the entire log with the inductive miner infrequent~\cite{Leemans.Robust_Process_Mining_with_Guarantees}.
The implementation and further results can be found online\footnote{\label{repo_URL}\url{https://github.com/fit-daniel-schuster/conformance_checking_for_trace_fragments}}.


Regarding the correctness of the proposed approaches: Baseline, Baseline~+ subsequent filtering and the Advanced approach, we compare the cost of the computed infix alignments. 
As the baseline approach considers all reachable markings as relevant, it is guaranteed that no other relevant markings exist.
Per trace infix, we find that all approaches yield infix alignments with identical costs.

\begin{figure}[tb]
    \centering
    \begin{subfigure}[b]{0.49\textwidth}
        \centering
        \includegraphics[clip,trim=0cm .25cm 0cm .25cm,width=\textwidth]{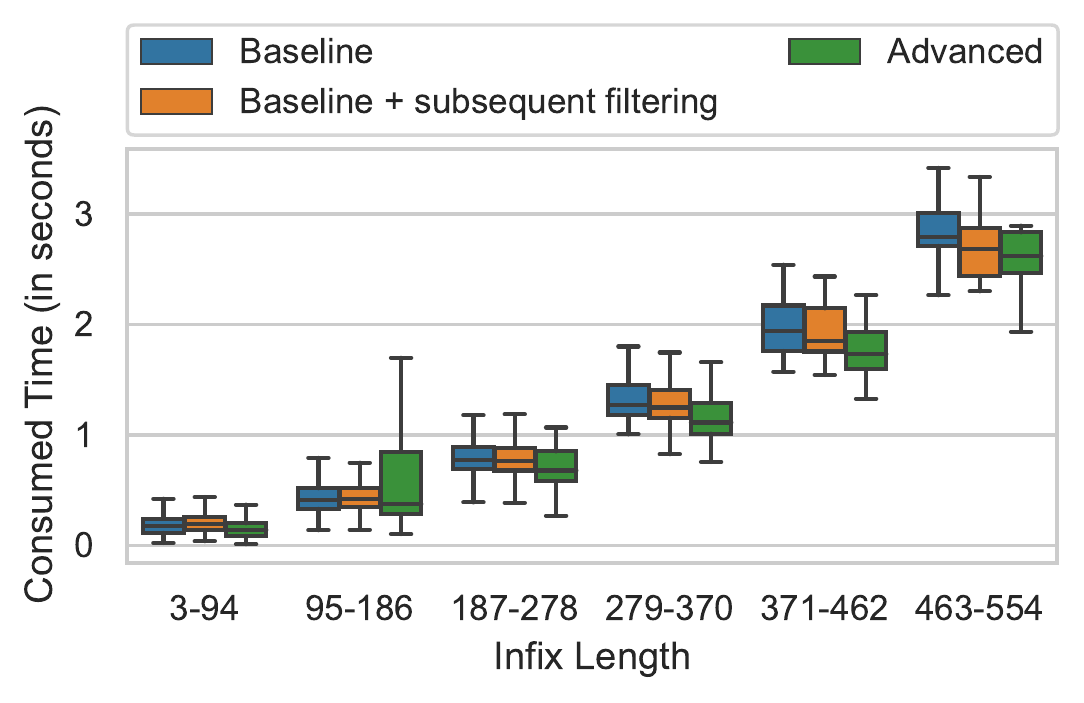}
        \caption{BPI Ch. 2019 event log}
        \label{fig:consumed_time_bpic_19}
    \end{subfigure}
    \hfill
    \begin{subfigure}[b]{0.49\textwidth}
        \centering
        \includegraphics[clip,trim=0cm .25cm 0cm .25cm,width=\textwidth]{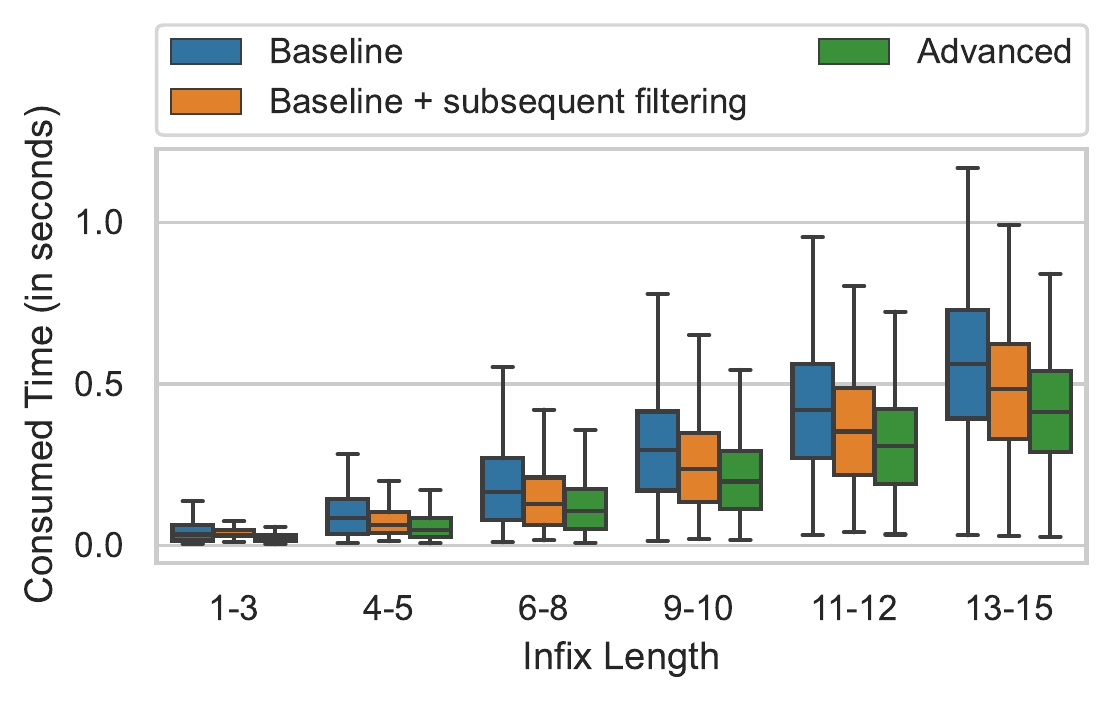}
        \caption{BPI Ch. 2020  event log}
        \label{fig:consumed_time_bpic_20}
    \end{subfigure}

    \caption{Time spent for computing infix alignments, i.e., Step 1-5 (cf. \autoref{sec:infix alignment-comp})}
    \label{fig:time_spent_alignment}
\end{figure}

\autoref{fig:time_spent_alignment} shows the overall time spent for the alignment computation, i.e., Step 1 to 5 (cf. \autoref{sec:infix alignment-comp}).
We find that using the advanced approach significantly shortens the overall alignment calculation time compared to the baseline approaches because the auxiliary WF-net produced by the advanced approach contains fewer silent transitions than the one created by the baseline approach.

\section{Conclusion}
\label{sec:conclusion}

This paper extended the widely used conformance checking technique alignments by defining infix and postfix alignments.
We presented two approaches for computing them, i.e., a baseline approach and an advanced approach assuming process trees as a reference model.
Our results indicate that the advanced approach outperforms the baseline if the reference process model is block-structured.

%
%
%
%
%
%
\bibliographystyle{splncs04}
\bibliography{main}

\begin{thebibliography}{10}
\providecommand{\url}[1]{\texttt{#1}}
\providecommand{\urlprefix}{URL }
\providecommand{\doi}[1]{https://doi.org/#1}

\bibitem{Adriansyah.phd}
Adriansyah, A.A.: {Aligning observed and modeled behavior}. Ph.D. thesis (2014)

\bibitem{conformance_checking_book}
Carmona, J., {van Dongen}, B., Solti, A., Weidlich, M.: {Conformance Checking}.
  {Springer} (2018)

\bibitem{Dunzer.conformance_checking_literature_review}
Dunzer, S., Stierle, M., Matzner, M., Baier, S.: {Conformance checking: A
  state-of-the-art literature review}. In: {Proceedings of the 11th
  International Conference on Subject-Oriented Business Process Management}.
  {ACM Press} (2019)

\bibitem{Leemans.Robust_Process_Mining_with_Guarantees}
Leemans, S.J.J.: {Robust Process Mining with Guarantees}. {Springer} (2022)

\bibitem{Schuster.Incremental_State-Space_Expansion}
Schuster, D., {van Zelst}, S.J.: {Online Process Monitoring Using Incremental
  State-Space Expansion: An Exact Algorithm}. In: {Business Process
  Management}. {Springer} (2020)

\bibitem{Taymouri.A_Recursive_Paradigm_for_Aligning_Observed_Behavior}
Taymouri, F., Carmona, J.: {A Recursive Paradigm for Aligning Observed Behavior
  of Large Structured Process Models}. In: {Business Process Management}.
  {Springer} (2016)

\bibitem{vanderAalst.The_Application_of_Petri_Nets_to_Workflow_Management}
{van der Aalst}, W.M.P.: {The Application of Petri Nets to Workflow
  Management}. {Journal of Circuits, Systems and Computers}  (1998)

\bibitem{replaying_history_on_process_models}
{van der Aalst}, W.M.P., Adriansyah, A., {van Dongen}, B.: {Replaying history
  on process models for conformance checking and performance analysis}. {WIREs
  Data Mining and Knowledge Discovery}  (2012)

\bibitem{soundness_wf_nets}
{van der Aalst}, W.M.P., {van Hee}, K.M., ter Hofstede, A.H.M., Sidorova, N.,
  Verbeek, H.M.W., Voorhoeve, M., Wynn, M.T.: {Soundness of workflow nets:
  classification, decidability, and analysis}. {Formal Aspects of Computing}
  (2011)

\bibitem{vanDongen.Aligning_Modeled_and_Observed_Behavior_A_Compromise_Between_Computation_Complexity_and_Quality}
{van Dongen}, B., Carmona, J., Chatain, T., Taymouri, F.: {Aligning Modeled and
  Observed Behavior: A Compromise Between Computation Complexity and Quality}.
  In: {Advanced Information Systems Engineering}. {Springer} (2017)

\bibitem{vanDongen.Efficiently_Computing_Alignments}
{van Dongen}, B.F.: {Efficiently Computing Alignments}. In: {Business Process
  Management}. {Springer} (2018)

\end{thebibliography}

\end{document}